\documentclass[useAMS,usenatbib]{mn2e}
\usepackage{graphicx} 
\usepackage{amssymb}
\newcommand {\hi} {{\rm H}\,{\small\rm I}}
\newcommand {\hicap} {{\rm H}\,{\scriptsize\rm I}}

\newcommand {\ovi} {{\rm O}\,{\small\rm VI}}
\newcommand {\ovicap} {{\rm O}\,{\scriptsize\rm VI}}
\newcommand {\ovii} {{\rm O}\,{\small\rm VII}}
\newcommand {\siii} {{\rm Si}\,{\small\rm II}}
\newcommand {\siiii} {{\rm Si}\,{\small\rm III}}
\newcommand {\siiv} {{\rm Si}\,{\small\rm IV}}
\newcommand {\cii} {{\rm C}\,{\small\rm II}}
\newcommand {\ciii} {{\rm C}\,{\small\rm III}}
\newcommand {\civ} {{\rm C}\,{\small\rm IV}}
\newcommand {\kms} {\,{\rm km\,s}^{-1}}

\newcommand {\pc} {\,{\rm pc}}

\newcommand {\kpc} {\,{\rm kpc}}

\newcommand {\cmmq}{\,{\rm cm^{-2}}}

\newcommand {\de}{^{\circ}}

\newcommand {\msun}{\,{\rm M}_\odot}

\newcommand{\zsun}{\,{Z}_\odot}

\newcommand{\Myr}{\,{\rm Myr}}
\newcommand{\Gyr}{\,{\rm Gyr}}
\newcommand{\K}{\,{\rm K}}

\newcommand {\msunyr}{\,{{\rm M}_\odot\,\rm yr}^{-1}}
\newcommand {\vlos}{v_{\rm LOS}}
\newcommand {\vlsr}{v_{\rm LSR}}

\newcommand{\avg}[1]{\left< #1 \right>} 

\title[On the origin of the warm-hot absorbers in the Milky Way's halo] 
{On the origin of the warm-hot absorbers in the Milky Way's halo}
\author[A. Marasco, F. Marinacci, \& F. Fraternali]
{
	A. Marasco$^{1}$\thanks{E-mail:antonino.marasco2@unibo.it},
	F. Marinacci$^{2, 3}$
	and F. Fraternali$^{1, 4}$
	\\
	$^{1}$Department of Physics \& Astronomy, University of Bologna, via Berti Pichat 6/b, 40127, Bologna, Italy\\
	$^{2}$Heidelberger Institut f\"{u}r Theoretische Studien, Schloss-Wolfsbrunnenweg 35, 69118 Heidelberg, Germany\\
	$^{3}$Zentrum f\"{u}r Astronomie der Universit\"{a}t Heidelberg, Astronomisches Recheninstitut, M\"{o}nchhofstr. 12-14, 69120 Heidelberg, Germany\\
	$^{4}$Kapteyn Astronomical Institute,  Postbus 800, 9700 AV, Groningen, The Netherlands\\
}
\begin{document}

\date{Accepted xxx Received xxx; in original form xxx}

\pagerange{\pageref{firstpage}--\pageref{lastpage}} \pubyear{2012}
\maketitle
\label{firstpage}

\begin{abstract}
Disc galaxies like the Milky Way are expected to be surrounded by massive coronae of hot plasma that may contain a significant fraction of the so-called missing baryons.
We investigate whether the local ($|\vlsr|\!<\!400\kms$) warm-hot absorption features observed towards extra-Galactic sources or halo stars are consistent with being produced by the cooling of the Milky Way's corona.
In our scheme, cooling occurs at the interface between the disc and the corona and it is triggered by \emph{positive} supernova feedback.
We combine hydrodynamical simulations with a dynamical 3D model of the galactic fountain to predict the all-sky distribution of this cooling material, and we compare it with the observed distribution of detections for different `warm' (\siiii, \siiv, \cii, \civ) and `hot' (\ovi) ionised species.
The model reproduces the position-velocity distribution and the column densities of the vast majority of warm absorbers and about half of \ovi\ absorbers.
We conclude that the warm-hot gas responsible for most of the detections lies within a few kiloparsecs from the Galactic plane, where high-metallicity material from the disc mixes efficiently with the hot corona. 
This process provides an accretion of a few $\msunyr$ of fresh gas that can easily feed the star formation in the disc of the Galaxy.
The remaining \ovi\ detections are likely to be a different population of absorbers, located in the outskirts of the Galactic corona and/or in the circumgalactic medium of nearby galaxies. 
\end{abstract}


\section{Introduction}
It is widely accepted that the baryonic content of the Universe, as estimated separately by the theory of primordial nucleosynthesis \citep{Pagel97} and studies on the anisotropy of the cosmic microwave background \citep{Spergel+07}, exceeds by a factor of $\sim10$ the amount of visible matter observed in galaxies.
Most of the baryons are therefore `missing', and are supposed to permeate the intergalactic space as a warm-hot intergalactic medium (WHIM).
In this framework, disc galaxies like the Milky Way should be embedded in massive structures of pressure-supported plasma at the virial temperature ($\gtrsim10^6\K$) called \emph{coronae} \citep{FukugitaPeebles06}, extending up to hundreds of kpc from the galaxy centres.

The observational quest for these cosmological coronae is still ongoing.
The X-ray surface brightness of these structures associated to Milky Way-like galaxies is considered too faint to be detected with the current generation of instruments \citep{Bregman07}.
Coronae are expected to be more massive in larger galaxies: in the giant spirals NGC 1961 and UGC 12591 X-ray emission has been detected at more than $50\kpc$ from the center \citep{AndersonBregman11,Dai+12}, indicating the presence of extended structures of hot gas accounting for $10-30\%$ of the missing baryons associated to those galaxies.
These coronae are expected to be metal-poor. 
\citet{HodgesBregman12} modelled the diffuse X-ray emission around NGC 891, and found the best fit by using a single gas component with a metallicity of $\sim0.1\zsun$, in agreement with the theoretical expectation.
In the Milky Way, the existence of a hot corona was originally hypothesized by \citet{Spitzer56} as a medium to provide pressure confinement to the High-Velocity Clouds \citep[HVCs,][]{WakkerVanWoerden97}.
\ovii\ absorption lines towards extragalactic sources, probing material at temperature of $\sim10^6\K$, failed to provide definitive evidence for an extended hot medium around the Galaxy \citep{Bregman07b,Yao+08}.  
Nevertheless, a number of indirect probes exist other than Spitzer's argument: the head-tail structure of several HVCs \citep{Putman+11}, the asymmetry of the Magellanic Stream \citep{Mastropietro+05} and the spatial segregation between gas-rich and gas-depleted satellites in the Local Group \citep[e.g.][]{GrcevichPutman09} may all be explained by ram-pressure exerted by the coronal medium.

Cosmological coronae can constitute an almost infinite reservoir of gas for feeding star formation in disc galaxies.
The Milky Way would completely deplete its current stock of gas in a few Gyr by forming stars at the current rate of $\sim3\msunyr$ \citep{Diehl+06}.
Instead, the star formation rate (SFR) of our Galaxy has decreased only by a factor $2-3$ over the last $10\Gyr$ \citep[e.g.][]{Twarog80,AumerBinney09}, which necessarily implies that its interstellar medium gets continuously replenished \citep[e.g.][]{FraternaliTomassetti12}.
Gas accretion of metal-poor material is also requested by chemical evolution models and to solve the so-called G-dwarfs problem \citep{Chiappini+97}.
For decades, HVCs have been considered as the main candidates for accreting low-metallicity material onto the Galaxy, but the most recent estimates of their accretion rate gives a value of $\sim0.08\msunyr$ \citep{Putman+12}, clearly insufficient to match the Galactic star formation rate.
The situation is similar in nearby galaxies where both anomalous clouds and minor mergers seem to account only for $\sim10\%$ of the accretion required \citep{Sancisi+08,Fraternali10}.
Hence, accretion from cosmological coronae seems to be the only viable possibility. 

The question is how the corona can cool and accrete onto the host galaxy.
It has been argued that cooling can occur spontaneously via thermal instability, producing ubiquitous clumps of cold gas in the coronal medium \citep[e.g.][]{MallerBullock04,Kaufmann06}. 
However, analytical calculations suggest that, in a corona stratified in a galactic potential, the combination of buoyancy and thermal conduction can efficiently suppress the growth of thermal perturbations \citep{Binney+09,Nipoti10,NipotiPosti12}.
Also, deep \hi\ surveys of galaxy groups systematically failed in finding isolated floating clouds with \hi\ masses $\gtrsim10^6\msun$ without optical counterparts at large distances from the host galaxies \citep[e.g.][]{Pisano+07,Chynoweth+09}. 
Finally, adaptive mesh refinement cosmological simulations of Milky Way-like galaxies do not show spontaneous cooling of the corona \citep{Joung+12}, and this result is now also confirmed by recent smoothed particle hydrodynamical simulations that adopt a more accurate treatment of the phase mixing \citep{Hobbs+12}.

Despite the above arguments, there is strong evidence for material cooling from the virial temperature inside the Galactic halo.
\ovi\ absorption lines towards extragalactic sources have been observed by \citet{Sembach+03} and \citet{Savage+03}, showing that material with temperatures of $2-5\times10^5\K$  is widespread in the sky.
The distances of the absorbers are unknown, but the low standard-of-rest velocities of the lines ($|\vlos|\!<\!400\kms$) suggest that this material is cooling from the Galactic corona.
More recently, lines of different ions (eg. \siii, \siiii, \siiv, \cii, \ciii, \civ), probing material in the $\sim2-20\times10^4\K$ range (assuming collisional ionisation equilibrium), have been observed in absorption towards extragalactic sources \citep{Shull+09, LH11, Lehner+12}.
In a few cases these features are detected in the spectra of Galactic halo stars, allowing us to place an upper limit to their distances of $\lesssim10\kpc$ from the disc \citep{Lehner+12}. 
After converting the measured ion column densities into a total gas mass, both \citet{Shull+09} and \citet{LH11} inferred an accretion rate onto the disc of $\sim1\msunyr$.

Sensitive \hi\ observations revealed that $\sim5-10\%$ of the \hi\ content of nearby disc galaxies is located a few kpc above the disc, forming the so-called `\hi\ halo' \citep[e.g.][]{Swaters+97}.
The kinematics of these structures is peculiar: they rotate with a lower speed with respect to that of the disc \citep{Oosterloo+07} and they show a general inflow motion \citep{Fraternali+02}.
In the Milky Way, \citet{MarascoFraternali11} found an \hi\ halo with a mass of $\sim3\times10^8\msun$ extended up to $2\!-\!3\kpc$ above the Galactic disc and inflowing towards the disc with a speed of $\sim35\kms$.
These \hi\ haloes are mainly built up by gas ejected from the disc via stellar feedback \citep{FB06,Heald06}.
The material ejected from the disc travels through the halo for $\sim80\!-\!100\Myr$ before returning back, creating a cycle of gas that goes under the name of `galactic fountain' \citep{ShapiroField76, Bregman80}. 
\citet{FB06,FB08} built a dynamical model of galactic fountain and applied it to the \hi\ observations in the haloes of two nearby galaxies: NGC 891 and NGC 2403.
They found that, while the models could easily reproduce the spatial distribution of neutral gas above the disc, the halo kinematics could be explained only by assuming that the fountain clouds accrete material with low angular momentum from an ambient medium.
\citet[][this latter hereafter M11]{Marinacci+10a,Marinacci+11} used 2D hydrodinamical simulations to show that this accretion can occur when fountain clouds interact with the corona: the latter cools via condensation in the turbulent wakes lagging behind the clouds.
This condensation is essentially produced by the mixing of coronal gas with metal-rich material coming from the disc.
\citet[][hereafter MFB12]{MFB12} used the dynamical model of \citet{FB08} including condensation from the corona to reproduce the \hi\ halo of the Milky Way. They found the best match for a current accretion rate of coronal gas onto the disc of $\sim2\msunyr$.
Further details on the works of M11 and MFB12 are reported in Section \ref{framework}.

\begin{figure*}
\begin{center}
\setlength\fboxsep{0pt}
\setlength\fboxrule{0.5pt}
\fbox{\includegraphics[width=\textwidth]{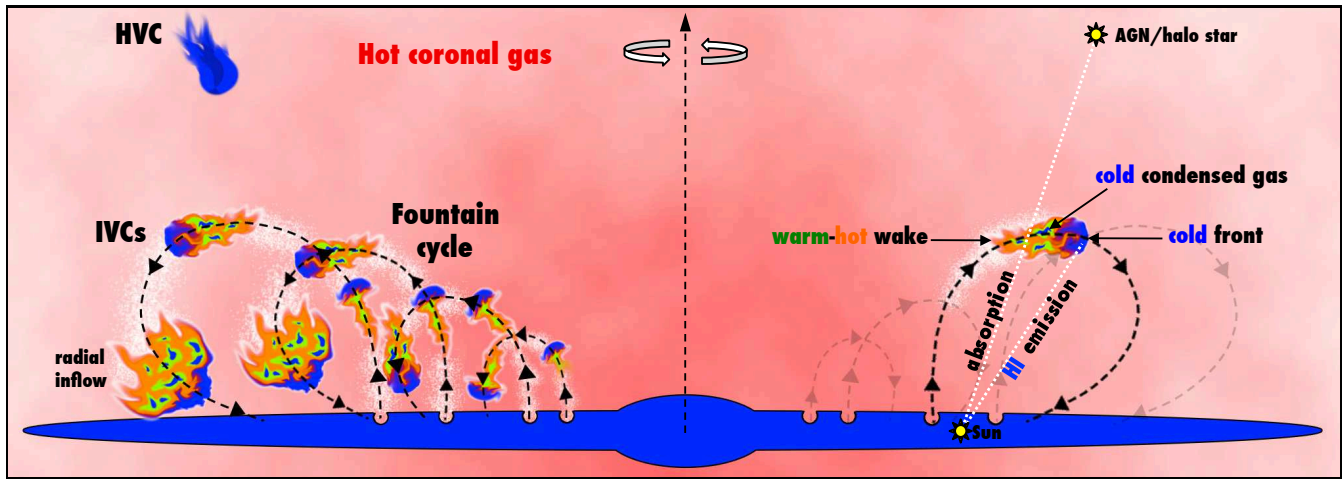}}
\caption{Scheme of the framework considered in this work. Galactic fountain clouds are ejected from the disc and travel through the halo. The azimuthal motion of the clouds is not shown. Hydrodynamical interactions between clouds and coronal material trigger the cooling of the latter inside the warm-hot turbulent wakes.
This fountain cycle produces a multiphase medium located within a few kiloparsecs above the disc. 
An observer looking towards a background source and intercepting a warm-hot wake detects absorption lines of ionised material.
An observer looking towards a cold front detects \hicap\ emission from an Intermediate-Velocity Cloud.}
\label{scheme}
\end{center}
\end{figure*}

Coronal gas cooling in the wake of the fountain clouds is expected to manifest in a wide range of temperatures below the virial one ($\sim1\!-\!2\times10^6\K$).
In this work we investigate whether this cooling material is responsible for the local ($|\vlos|\!<\!400\kms$) warm-hot absorptions of different ions (\ovi, \siiii, \siiv, \cii, \civ) observed toward AGNs/QSOs and distant halo stars in the works of \citet{Lehner+12}, \citet{Sembach+03} and \citet{Savage+03}.
For this purpose we make use of the dynamical model of MFB12, but instead of studying the neutral hydrogen component of the cloud+wake systems, we focus on material at higher temperatures.
In Section \ref{model} we analyse the simulations of M11 in order to include the dynamics of the warm-hot gas in the MFB12 model.
In Section \ref{results} we compare the predictions of the resulting models with the available data of warm-hot absorptions.
In Section \ref{discussion} we discuss our results and in Section \ref{conclusions} we draw our conclusions.

\section{The model}\label{model}
Here we describe the scheme of fountain-clouds -- corona interaction adopted throughout the paper. 
Our framework is set up mainly by the works of M11 and MFB12, whose results are summarised in the following.

\subsection{Setting up the framework}\label{framework}
M11 investigated the interaction between fountain clouds and the cosmological corona by simulating clouds of cold ($10^4\K$) gas at solar metallicity traveling through a homogeneous medium with a temperature of $2\times10^6\K$ \citep[see][]{FukugitaPeebles06} and a metallicity of $0.1\zsun$ \citep[see][]{HodgesBregman12}.
They found that, as the cloud moves through the coronal medium, Kelvin-Helmholtz instability develops. 
This produces a turbulent wake behind the cold gas in which material stripped from the cloud mixes with the ambient hot gas, increasing the metallicity of the latter and thus decreasing its cooling time. 
As a result, the amount of warm and cold gas in the cloud's wake increases exponentially with time, to the detriment of the coronal gas: i.e. the galactic fountain triggers the cooling of the cosmological corona.
Moreover, as the cloud moves through the corona, the cold and the hot phases exchange momentum with each other.
The efficiency of this mass and momentum transfer critically depends on the relative velocity between the two gas phases.
Being at least partially pressure supported, the Galactic corona cannot corotate with the cold gas. 
M11 found that the corona ceases to acquire momentum when the relative velocity between the two phases drops below $\sim75\kms$, because condensation immediately recaptures the momentum gained by the hot phase.
The existence of this velocity threshold suggests that the corona must spin, at least in the regions closer to the Galaxy, but with a velocity $75\!-\!100\kms$ lower than that of the disc.

Motivated by these results, MFB12 modelled the \hi\ halo of the Milky Way by using a dynamical axi-symmetric 3D model of galactic fountains where cloud particles, ejected from the disc at a rate proportional to the local star formation rate density, travel through the Galactic halo before returning back to the disc.
The particles are subjected to the gravitational force of the Galaxy and the hydrodynamical forces due to the interaction with the corona.
The latter rotates with a speed $75\kms$ lower than the local circular speed, consistently with the results of M11.
The particles accrete mass from this spinning medium at a rate consistent with the condensation rate of coronal gas in the wakes of the fountain clouds.
Thus, these particles are in fact representative of the cloud+wake system.
The model has three free parameters: the ejection velocity of the particles, the fraction of the orbit during which particles are not visible in \hi\ because hydrogen is ionised, and the rate at which clouds are accreting material from the corona.
The best parameters are derived by comparing the simulated data, produced for an observer placed at $R\!=\!R_\odot$, with the \hi\ datacube of the \emph{Leiden-Argentine-Bonn} (LAB) survey \citep{Kalberla+05}.
Models including condensation perform much better than ballistic fountain models where the interaction between clouds and corona is neglected.
The condensation of coronal gas provides an accretion of pristine material onto the disc at a rate of $\sim2\msunyr$, remarkably similar to the star formation rate of the Milky Way.
The accretion rate remains of a few $\msunyr$ for rotational lags ranging from $50\kms$ to $100\kms$ (MFB12).
The model reproduces the \hi\ emission of the classical Intermediate-Velocity Clouds (IVCs).
This implies that the Galactic \hi\ halo comprises thousands of IVC clouds, with the classical complexes being only a local component.

\begin{figure*}
\begin{center}
\includegraphics[width=0.9\textwidth]{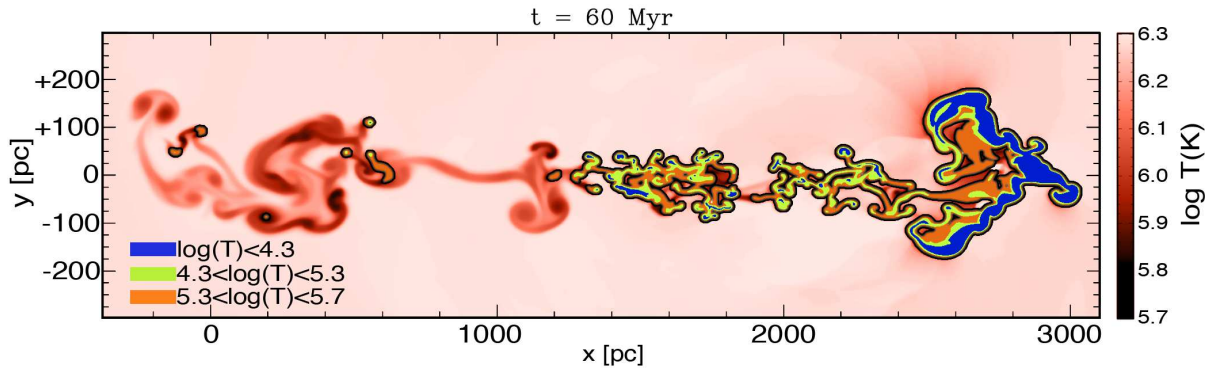}
\caption{Snapshot at time $t\!=\!60\Myr$ from the hydrodynamical simulation of M11. The initial relative velocity between the cloud and the hot environment is $75\kms$, and the initial mass of the cloud is $2.4\times10^4\msun$. The colours code the different ranges of temperature defined in Section \ref{wake}: \emph{blue} regions represent the cold gas ($\log(T)\!<\!4.3$), \emph{green} regions represent the warm gas ($4.3\!<\!\log(T)\!<\!5.3$) and \emph{orange} regions represent the hot gas ($5.3\!<\!\log(T)\!<\!5.7$). The black contour is at $log(T)\!=\!5.7$. \emph{Red} regions represent material at higher temperatures.}
\label{wake_hydro}
\end{center}
\end{figure*}

\begin{figure}
\begin{center}
\includegraphics[width=0.4\textwidth]{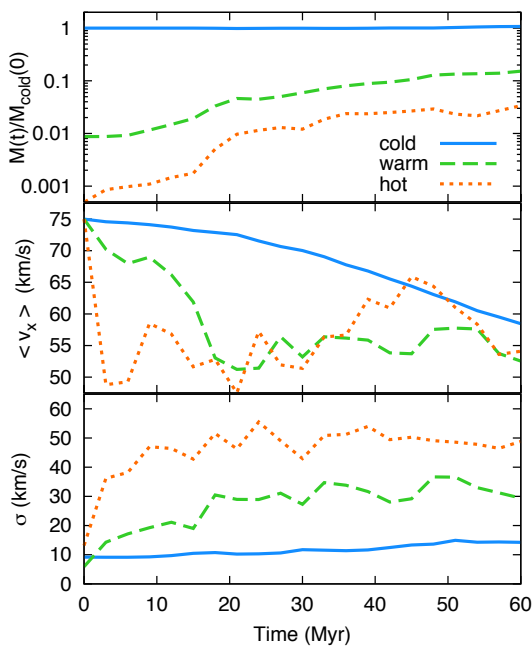}
\caption{Time-evolution of different quantities in the hydrodynamical simulation of M11. Different lines show gas in different ranges of temperature, as indicated in the top panel (see also Section \ref{wake}). \emph{Top panel:} mass ratio between gas at a given temperature and the cold one at $t\!=\!0$; \emph{middle panel:} mass-weighted velocity along the x-axis; \emph{bottom panel:} mass-weighted total (turbulent+thermal) velocity dispersion along the x-axis.}
\label{analyze_hydro}
\end{center}
\end{figure}

The global picture that emerges from these works is sketched in Fig.\,\ref{scheme}.
The gaseous halo of the Galaxy is populated by high metallicity fountain clouds (IVC-like) that interact with the metal poor coronal gas. 
This latter is entrained by the clouds' wakes and cools efficiently.
The system fountain cloud+wake is multiphase (see Section \ref{wake}).
MFB12 have shown that the cold phase of this medium can account for \hi\ observations in the Galactic halo. 
In this work, instead, we compare the warm-hot phase with the detections of different absorption lines of ionised material.

\subsection{The wake of a fountain cloud}\label{wake}
We use the simulations of M11 to quantify how the mass and the velocity of a fountain-cloud wake evolve with time for different ranges of temperature.
The simulations include radiative cooling, and the relative velocity between the cold and the hot phase is fixed to $75\kms$ for the reasons discussed in Section \ref{framework}. The effects of different relative velocities are discussed in Section \ref{discussion}.
We assume collisional ionisation equilibrium (CIE) to infer the ranges of temperature in which the considered ions are supposed to occur \citep{SutherlandDopita93,Bregman07}.
We distinguish three ranges of temperature: 
\begin{itemize}
\item \emph{cold gas}: gas with $\log{T}\!<\!4.3$, mainly \hi; 
\item \emph{warm gas}: gas with $4.3\!<\!\log{T}\!<\!5.3$, responsible for the absorptions in \siiii,\,\siiv,\,\cii,\,\ciii\,and \civ;
\item \emph{hot gas}: gas with $5.3\!<\!\log{T}\!<\!5.7$, responsible for the absorptions in \ovi.
\end{itemize}
With the term \emph{corona} instead we indicate the low density environment at the virial temperature ($\gtrsim\!10^6\K$) that surrounds the Galaxy. 
Here, we assume that the ion abundances of a given atomic species in the gaseous halo depend only on the temperature of the medium, thus photoionisation from stellar disc and extragalactic sources is neglected.
The influence of photoionisation on the \ovi\ abundances should not be relevant \citep[see][]{Sembach+03}, while it may have an impact on the other species considered (see Section \ref{refinement}).
Fig.\,\ref{wake_hydro} shows a snapshot at time $t\!=\!60\Myr$ taken from the simulation analysed. 
The different ranges of temperature considered are coded by three different colours. 
Clearly, the cold gas concentrates in the rightmost part of the system, but a few knots are present also in the wake due to coronal material cooling down to $\sim10^4\K$. 
The warm and the hot gas are present both immediately behind the cold front and in the wake, which extends about $2\kpc$.

For each snapshot of the simulation, we evaluate the mass $m(t)$, the mean velocity parallel to the motion of the cloud $\avg{v_x(t)}$ and the velocity dispersion $\sigma_{\rm turb}(t)$ for the gas in the above ranges of temperature.
The latter, evaluated as the standard deviation of the $v_{x}$ distribution, is assumed to be isotropic.
The total one-dimensional velocity dispersion $\sigma(t)$ is computed as the quadratic sum of the turbulent and the thermal component:
\begin{equation}\label{sigma}
\sigma(t) = \sqrt{\sigma_{\rm turb}^2(t) + \frac{k_B T(t)}{m_{\rm ion}}}
\end{equation}
where $T(t)$ is the average temperature of the gas at a given time.
Note that the thermal component depends on the mass of the ion considered. 
Fig.\,\ref{analyze_hydro} shows the time evolution of the above-mentioned quantities.
The panel on top clearly shows that the mass of the warm and the hot gas, negligible at the beginning of the simulation, increases by more than one order of magnitude in $60\Myr$, while the mass of the cold gas increases only by a few percents with respect to the initial value \citep[see also][]{Marinacci+10a}.
The velocity of the cold gas centroid smoothly decreases with time because of the combined effects of drag and condensation (see MFB12).
The velocities of the warm and the hot components remain below that of the cold material for the whole simulation,  showing an irregular trend with time.
As a consequence, the warm and the hot material falls further and further behind the cold front during the cloud's orbit.
The total velocity dispersion, after the first $20\Myr$, flattens to the values of $12\kms$ for the cold gas, $31\kms$ for the warm gas and $49\kms$ for the hot gas.
Since the thermal velocity dispersion is only $6\kms$ for the warm gas (considering Carbon) and $14\kms$ for the hot gas (considering Oxygen), the turbulent component always dominates over the thermal one in these cases.
For the cold gas the two components are comparable. 
The warm-hot gas in the fountain-cloud wakes has sub-solar metallicity, being it produced by the mixing between a metal-rich component, i.e. the fountain cloud, and the metal-poor coronal gas. 
In the simulation analysed, the mass-weighted metallicity is $\sim0.5\zsun$.

\subsection{From simulations to `modelcubes'}\label{simtocube}
In the previous Section we showed how the wake of a single fountain cloud develops and evolves with time.
However, the halo of the Milky Way is supposed to be populated by thousands of these objects \citep{MarascoFraternali11}, so we need a global 3D model to determine how the warm-hot gas of the fountain-cloud wakes distributes in the sky.
We take the axi-symmetric model built by MFB12 to reproduce the cold (\hi) phase of the halo and we include the warm and the hot phases as follows.

We associate to each `cold particle' that is ejected from the disc a `warm-hot particle' that follows the orbit of the former.
The cold particles represent the \hi\ phase of the system cloud+wake, whose time evolution is already fixed in the model by the fit to the LAB \hi\ survey (MFB12).
The mass and the velocity of the warm-hot particles are instead set accordingly to the trends shown in Fig.\,\ref{analyze_hydro}.
Specifically, we focus on the mass ratio between the warm (or hot) and the cold gas, and on the velocity gap between them. 
We fit the first with a power law and the second with a linear function.
We exclude from the fit all the points at $t\!<\!25\Myr$, because the simulations begin with an unrealistic spherical cloud (see also Section 5.1 of MFB12). 
In the dynamical model, at each timestep we set the masses of the warm-hot particles by multiplying the mass of the cold particle - which is known in the MFB12 model - by the mass ratio derived above, while the kinematics is obtained by subtracting the velocity gap from the velocity of the cold particle.

The dynamical model of MFB12 has a number of stochastic variables, like the ejection angles and velocities of the particles, which can produce fluctuations in the final synthetic datacube. 
To reduce their impact, we built an average synthetic datacube for the warm (and the hot) gas by combining the cubes resulting from $15$ different runs.
We applied a gaussian smoothing in the velocity direction to the average cube accordingly to the one-dimensional velocity dispersion associated to the warm and hot components (see Section \ref{wake}).
Finally, a spatial smoothing of $5\de$ is applied in order to wash out the clumpyness due to the discrete nature of our model.  
The resulting synthetic datacube - or, simply, the `modelcube' - contains the mass of the warm (or hot) gas as a function of sky position and line-of-sight velocity in the local standard of rest.
We indicate with $I(l,b,v)$ the column density of the warm/hot gas per velocity unit at the position $(l,b)$ (typical units in $\msun\pc^{-2}/\kms$). 
We stress that this modelcube is not derived by fitting our model to the effective distribution of warm-hot gas in the Milky Way's halo, but it fully relies on our physical understanding of both the halo dynamics and the disc-corona interaction.
The only fit involved in the process is that of the parameters of the galactic fountain, which has been performed by MFB12 using the LAB \hi\ survey.

\subsection{Comparison with the data}\label{dataanalysis}
In the literature there are several datasets reporting detections of absorption features towards AGNs and, in some cases, Galactic stars \citep[e.g.][]{Zsargo+03,Bowen+08}.
Here we focused on three datasets that provide accurate measurements of the velocity centroids of the absorption lines, which are fundamental for our analysis (see below), and consider a large amount of background targets. 
These data are published in: \citet[][warm absorbers with $|\vlos|\!>\!90\kms$]{Lehner+12}, \citet[][\ovi\ absorbers with $|\vlos|\!>\!90\kms$] {Sembach+03} and \citet[][\ovi\ absorbers with $|\vlos|\!<\!90\kms$]{Savage+03}.
Each dataset is a list of observed lines of sight in the direction of several AGNs/QSOs and, in a few cases, halo stars.
For each of them, if one or more absorptions are detected, the local standard of rest velocity centroids of each absorption line is reported.
The \ovi\ data contain also information on the line-width and on the \ovi\ column density of the detections, or an upper limit to the column density in the case of non-detections.
All together, 84 warm detections and 175 \ovi\ detections are present in the datasets analysed.
Using the approach that is discussed in Section \ref{missingdetections} we identify 12 warm detections and 16 \ovi\ detections that are likely to be related to the ionised gas surrounding the Magellanic Clouds/Stream \citep{BlandHawthorn+07}. 
These absorptions are not considered in our analysis, thus the effective number of detections decreases to 72 for the warm component and to 159 for the \ovi\ one.
In any case, the inclusion of these features does not change our results significantly.

In order to compare our model with the data we must realise that the former represents an average, smooth, and axi-symmetric realization of the warm-hot gas distribution resulting from the fountain mechanism.
This is because the number of particles that are ejected in our model is much larger than the number of clouds that effectively populate the Galactic halo.
Hence, we can not compare our model with the details of the single detection, which depend on the chance of intercepting the wake of a specific fountain cloud.
Instead, we adopt the following statistical approach.
In our model, the intensity of the generic pixel $I(l,b,v)$ is proportional to the probability of finding an absorption at the position $(l,b)$ in the sky and line-of-sight velocity $v$.
We define the quantity
\begin{equation}\label{omega}
f(I_i) = \frac{F(I_i)}{F(0)}
\end{equation}
where
\begin{equation}
F(I_i) = \Delta l\, \Delta b\, \Delta v\, \sum_{l,b,v} I(>\!I_i) (l,b,v) \cos(b)\,.
\end{equation}
In the equation above, $ \Delta l$, $\Delta b$ and $\Delta v$ are the pixel separations in the three directions, the factor $\cos(b)$ stands for the projection effects and the sum is extended to all the pixels with intensity larger than $I_i$.
Note that since $F(0)$ is the total flux of the modelcube, we have $0\!\le\!f\!\le\!1$.
Hence a fraction $f(I_i)$ of the total flux is contained in pixels with $I>I_i$.
Using eq.\,(\ref{omega}) we define three confidence levels $I_1$, $I_2$ and $I_3$ as follows:
\begin{displaymath} 
	f(I_i) = \left\{ \begin{array}{ll}
	0.6827 & \textrm{if $i=1$}\\ 
	0.9545 & \textrm{if $i=2$}\\ 
	0.9973 & \textrm{if $i=3$}
	\end{array} \right. 
\end{displaymath}
The meaning of these values is straightforward: if an absorption detected in $(l,b,v)$ falls inside the first contour level (i.e. $I(l,b,v)\!>\!I_1$), then the model reproduces that detection with $1\sigma$ level of confidence (by analogy with a Gaussian statistics). 
If it falls inside the second one ($I_2\!<\!I(l,b,v)\!<\!I_1$), the level of confidence is $2\sigma$ and so on.

We use these confidence levels to compare our models with the data in two ways.
On the one hand, we show the confidence levels in a series of longitude-velocity ($l\!-\!v$) diagrams at different latitudes to visualize the probability of finding an absorption as a function of the position in the sky and the line-of-sight velocity.
We overlay the detections on these diagrams to give a visual comparison between the two.
On the other hand, we perform a statistical comparison between the predicted and the observed distribution of the absorptions in the $(l,b,v)$ space. 
We interpret the discrepancy between the two distributions with the presence of a \emph{different} population of absorbers, i.e. not produced by the fountain-cloud--corona interaction, that contaminates the data.
We implement an algorithm based on the Kolmogorov-Smirnov (KS) test to determine the fraction of detections that must be discarded in order to have the two distributions consistent with each other.
The algorithm is described in the next Section.

We point out that, with the method described, we can not predict the absolute probability that an absorption occurs at a given $(l,b,v)$.
Instead, our model gives the relative probability of an absorption to occur at a specific location $(l,b,v)$ rather than a different one.
Specifically, we discriminate the regions in the $(l,b,v)$-space where the warm-hot gas is present in a significant amount from those depleted of it, and we compare how the observed detections and the modelled warm-hot gas are distributed in that space (Section \ref{results}).
In Section \ref{coldens_comp} we show that the density of hot gas predicted by our model in the regions where absorptions are observed is remarkably consistent with the measured column density of the absorbers. 

\subsection{Iterative KS algorithm}\label{KS}
Given a modelcube of warm/hot gas and a list of $n$ warm/\ovi\ absorption features, we compute the fraction of detections that is consistent with the model using the following iterative procedure:
\begin{enumerate}
\item we evaluate the fraction and the cumulative fraction of detections observed and predicted as a function of the confidence level described in Section \ref{dataanalysis}; 
\item we use the numerical code of \citet{Marsaglia+03} to compute the KS critical value $K(n,d)$, where $d$ is the largest absolute distance between the two cumulative fractions; 
\item if $K(n,d)\!>\!0.95$, i.e. if the model is rejected with a probability larger than $95\%$, we remove from the data a single detection. We choose it randomly within the confidence level with the largest (positive) difference between the fraction of detections observed and predicted. 
\end{enumerate}
This procedure is iterated until the probability of rejection drops below $95\%$.
We consider the remaining detections consistent with the model.
We evaluate an error bar on the fraction of detection to discard by considering different thresholds for the KS critical value: the lower limit is chosen at $K(n,d)\!=\!0.99$, the upper limit at $K(n,d)\!=\!0.90$.
Note that in the calculation of the predicted fractions, only the observed lines of sight must be considered.
This approach allows us to correct for the fact that the background sources are not isotropically distributed in the sky.

\section{Results}\label{results}
\begin{figure*}
\includegraphics[height=20cm]{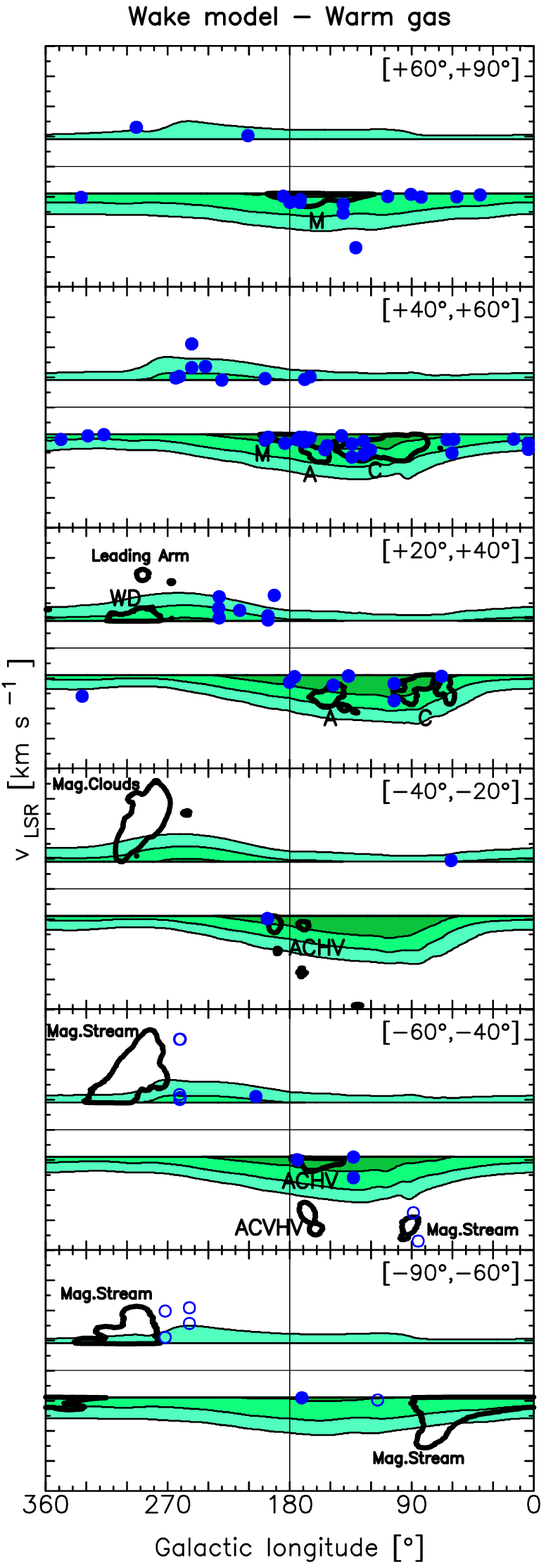}
\includegraphics[height=20cm]{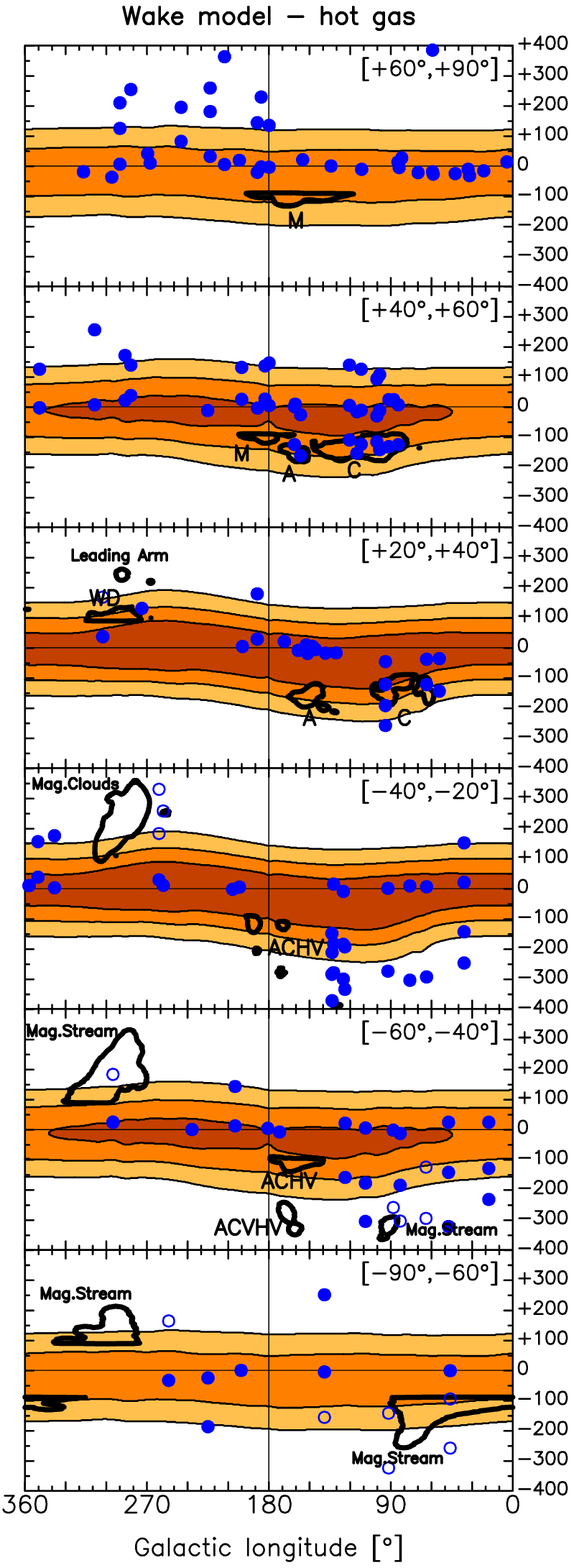}
\caption{Longitude-velocity diagrams in 6 different latitude bins (indicated at the top right corner of each panel) showing the confidence contours of our model of fountain wakes.
\emph{Left panels:} warm phase of the fountain wakes. Circles represent warm detections from \citet{Lehner+12}.
\emph{Right panels:} hot phase of the fountain wakes. Circles represent \ovicap\ detections from \citet{Sembach+03} and \citet{Savage+03}.
Empty circles represent detections considered related to the Magellanic Clouds/Stream.
Contour levels from the innermost to the outermost at $1\sigma$, $2\sigma$ and $3\sigma$ (see Section \ref{dataanalysis}). The \hicap\ emission of the classical HVCs is reported in the various panels as the thick contours, labelled with the name of the respective complexes.}
\label{wake_lv}
\end{figure*}

Fig.\,\ref{wake_lv} shows different longitude-velocity diagrams for our model of galactic fountain wakes. 
Each panel represents a given latitude bin.
Contour levels are derived with the approach described in Section \ref{dataanalysis}, and both the warm and the \ovi\ detections are overlayed on the diagrams.
Detections that we consider related to the Magellanic system are represented by empty circles (see Section \ref{missingdetections}).
The velocity range $|v_{\rm LSR}|\!<\!90\kms$ is excluded from the observations of \citet{Lehner+12}, so the probability of finding a warm absorption at these velocities is set to zero by blanking all the pixels with $|v_{\rm LSR}|\!<\!90\kms$ in the modelcube of warm gas.
Also, in both the modelcubes we blanked the pixels at $|b|\!<\!20\de$ because these latitudes are systematically excluded by the observations in order to avoid overlapping with the Galactic disc.
Since detections are not distributed isotropically in the sky, contours in Fig.\,\ref{wake_lv} are expected to `encompass' the data rather than matching their overall pattern.
Globally, the warm absorptions concentrate at velocities $|v_{\rm LSR}|\!<\!200\kms$ while the \ovi\ detections are much more spread over the whole velocity range.
This is consistent with our models, which predict a larger velocity dispersion for the hot gas in the fountain wakes (Fig.\,\ref{analyze_hydro}).
We have seen that this dispersion is almost completely due to turbulence (Section \ref{wake}), which in turn depends on the physical mechanism of production, i.e. the fountain-cloud--corona interaction. 
Fig.\,\ref{wake_lv} shows that most detections are inside $3\sigma$ level of confidence, especially in the case of the warm component, but the majority is not within $1\sigma$.
Our model predicts that most of the warm-hot gas is produced in the descending part of the fountain orbits because of the increase in mass that the particles experience in time (Fig.\,\ref{analyze_hydro}), which implies that they are more massive when they are falling back to the disc.
Since these massive infalling particles have $\vlos\!<\!0$ by definition, the confidence contours shown are systematically shifted towards the negative velocities.
Consistently, detections occur more frequently at negative than positive velocities.
However, note that the warm data at $+40\de\!\le\!b\!\le\!+60\de$ and the \ovi\ data at $+60\de\!\le\!b\!\le\!+90\de$ show an overabundance of detections at positive velocities with respect to the prediction of our model.
We discuss this point in Sections \ref{refinement} and \ref{missingdetections}.
In Fig.\,\ref{wake_lv} we also show the \hi\ emission of several classical High-Velocity complexes, which have been derived from the LAB survey smoothed at $5\de$ of resolution.
Several warm and hot detections at negative velocities overlap with the \hi\ emission of the complexes A, C and M.
The relation between the absorption features, the HVCs and the gas in the wakes of the fountain clouds is discussed in Section \ref{HVCs}.

\begin{figure}
\begin{center}
\includegraphics[width=0.5\textwidth]{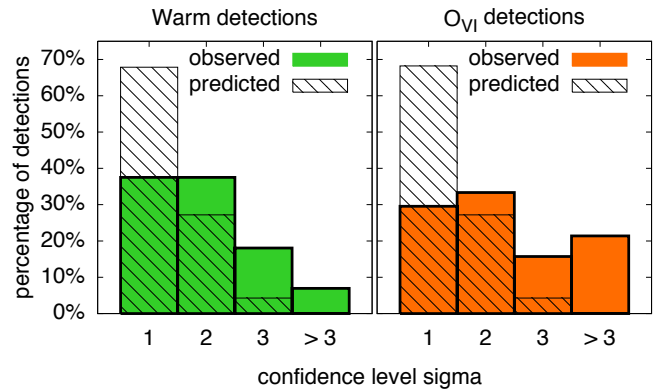}
\caption{Comparison between the fractions of detections observed (\emph{filled histograms}) and predicted (\emph{hatched histograms}) within a given confidence level by our model of fountain wakes. Warm absorptions are shown in the left panel, \ovicap\ in the right.}
\label{histo_1}
\end{center}
\end{figure}

Fig.\,\ref{histo_1} compares the fractions of absorptions detected within a given confidence level (filled histogram) with those predicted by our model of fountain wakes (hatched histogram).
The latter follows by definition a Gaussian distribution (see Section \ref{dataanalysis}) and are obtained by considering only the observed lines of sight, thus they are not affected by the non-isotropic distribution of the absorbers.
We use the algorithm described in Section \ref{KS} to derive the fraction of detections that our model of fountain wakes reproduces, obtaining $73.5^{+4.8}_{-2.2}\%$ for the warm case and $54.2^{+2.6}_{-1.3}\%$ for the \ovi\ case.
Thus, even if most of the detections are consistent with our scheme, there is a significant fraction that is not in agreement.
We discuss this further in Sections \ref{refinement} and \ref{missingdetections}.

\begin{figure*}
\includegraphics[width=\textwidth]{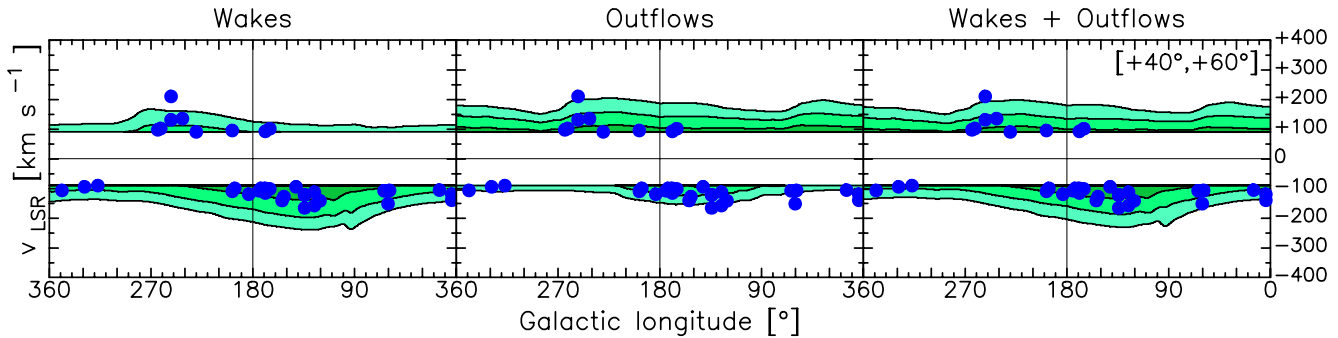}
\caption{Longitude-velocity diagram at $+40\de\!\le\!b\!\le\!+60\de$ for the warm gas component of our model of fountain wakes (\emph{left panel}), fountain outflows (\emph{middle panel}) and fountain wakes+outflows (\emph{right panel}). The points represent the warm detections from \citet{Lehner+12}. Contour levels from the innermost to the outermost at $1$, $2$ and $3\sigma$ (see Section \ref{dataanalysis}).}
\label{wake_outflow}
\end{figure*}

\subsection{Fountain wakes + outflows}\label{refinement}
The classical picture of the galactic fountain states that the gas escaping from the superbubble is ionised and it remains undetected in \hi\ until it cools and recombines.
This scenario is supported also by direct observations.
A classical case is that of the Ophiucus superbubble in the inner Galaxy, which shows large quantities of ionised gas visible in H$\alpha$ \citep{Pidopryhora+07}.
This ionised gas may largely contribute to the low and high-ion absorptions, thus it is important to include this material in our model.

Since large uncertainties are present in the physics of the gas escaping from a superbubble, a rigorous treatment of this component is extremely difficult and it would be beyond the purposes of this work.
Instead, we adopt a simple approach, based on the results found for the neutral gas.
MFB12 already investigated what fraction of the fountain orbits must remain `invisible' in order to reproduce the \hi\ emission of the Galactic halo.
They found that the recombination should occur when gravity reduces the vertical speed of the ejected cloud by $\sim30\%$, i.e. at about one-third of the ascending part of the orbit.
Before that time, the gas is ionised and therefore it is not visible in \hi.
Here we assume that this ionised outflowing material, starting from the moment of the ejection and until the recombination, is in a \emph{warm} phase and therefore it can be included in our modelcube of warm fountain gas.
We have already mentioned (Section \ref{wake}) that photoionisation from the disc can play an important role in shaping the distribution of the warm absorber material.
Including the ionised outflowing component of the galactic fountain corresponds to maximize the effect of the photoionisation in a region of few hundreds of parsecs above the Galactic disc.
In order to model the position-velocity distribution of these outflows, we use the best-fit model of MFB12 and consider only the first third of the ascending part of the fountain orbits.
Also in this case, the modelcube related to this material is obtained by averaging $15$ different runs and is smoothed to $5\de$ of resolution.
A critical ingredient to include in the model is the amount of velocity dispersion of the outflowing material, which depends on the turbulence of the gas escaping from the disc.
For simplicity we set this value to $31\kms$, the value derived for the warm material of the fountain wakes in the simulation analysed (Section \ref{wake}).
The resulting modelcube is then summed pixel by pixel to that obtained for the warm particles and is compared with the warm detections of \citet{Lehner+12} using the method described in Section \ref{dataanalysis}.

Fig.\,\ref{wake_outflow} shows the longitude-velocity diagram around $b=+50\de$ for our models of fountain wakes (on the left), fountain outflows (in the middle) and fountain wakes+outflows (on the right) for the warm component.
As expected, in the model of fountain outflows the gas occupies preferentially the positive velocities, where a consistent fraction of detections takes place.
However, outflows alone can not explain the whole set of detections.
The last model, obtained by combining the first two, provide a much better description of the space-velocity distribution of the warm dataset.

\begin{figure}
\begin{center}
\includegraphics[width=0.3\textwidth]{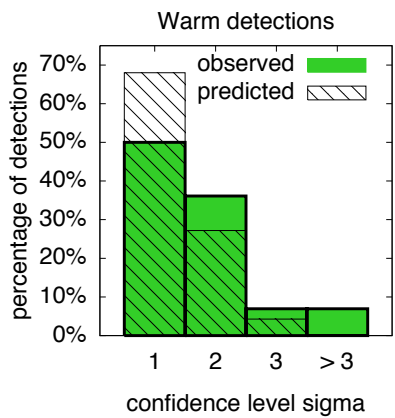}
\caption{As in Fig.\,\ref{histo_1}, but for a model of fountain wakes that include outflowing ionised gas. The comparison is with the warm detections from \citet{Lehner+12}}.
\label{histo_2}
\end{center}
\end{figure}

Fig.\,\ref{histo_2} compares the fractions of warm absorptions detected within a given confidence level with those predicted by our model of fountain wakes + outflows.
The fraction of detections reproduced by the latter, inferred with our KS-based algorithm, is $94.4^{+5.6}_{-2.5}\%$.
This fraction is significantly larger than that estimated for a model of fountain wakes alone ($73.5^{+4.8}_{-2.2}\%$, see Section \ref{results}), indicating that supernova-driven outflows from the disc play a significant role in populating the Galactic halo with ionised gas.

We tested whether our model of warm wakes with outflows is consistent with the Si ions absorption features found by \citet{Shull+09}. We found that only one-third of these detections are reproduced by our model, which suggests that the datasets of \citet{Shull+09} and \citet{Lehner+12} are not fully compatible with each other.
Further details on this analysis are given in \citet{Fraternali+13}.

\subsection{Column density comparison}\label{coldens_comp}
The amount of warm and hot gas present in the Galactic halo as a function of position and velocity is one of the most important predictions of our model.
It is crucial to compare this prediction with the column densities observed for the warm and hot detections.

We first focus on the hot component, which we compare with the \ovi\ column density measurements of \citet{Sembach+03} and \citet{Savage+03}.
At the position $(l,b)$, for a mean velocity $v$ and a line-width $b_w$, the \ovi\ column density predicted by our model can be evaluated as
\begin{equation}\label{NOVI}
N_{\rm O_{VI}} = \int_{v-2\sigma_w}^{v+2\sigma_w} \frac{n_{\rm O}}{n_{\rm H}}(Z)\,\frac{n_{\rm O_{VI}}}{n_{\rm O}}(T) \,I(l,b,v)\,dv
\end{equation}
where $\sigma_w\!=\!b_w/\sqrt2$ \citep{Sembach+03}, $\frac{n_{\rm O}}{n_{\rm H}}$ is the oxygen abundance, that depends on the metallicity $Z$ as follows:
\begin{equation}\label{Oxygen}
\log\left(\frac{n_{\rm O}}{n_{\rm H}}\right)(Z) = \log\left(\frac{n_{\rm O}}{n_{\rm H}}\right)_{\odot} + \log\left(\frac{Z}{Z_\odot}\right)  
\end{equation}
and $\frac{n_{\rm O_{VI}}}{n_{\rm O}}$ is the \ovi\ ion fraction, that for $\log(T)\!=\!5.45$ is $0.22$ if one assumes CIE (Sutherland \& Dopita 1993).
In eq.\,(\ref{Oxygen}) we assume a constant value for $\log(Z/Z_{\odot})$ of $-0.35$ obtained as the mass-weighted metallicity of the hot gas in the simulations, while we fix $\log(n_{\rm O}/n_{\rm H})_{\odot}$ at $-3.07$ accordingly to \citet{AndersGrevesse89}.
Note that we are assuming implicitly that the abundance ratio between different metals is solar-like.
In the case of non-detections, we still use eq.\,(\ref{NOVI}) to give an upper limit to the \ovi\ column density. 
In this case, the integral is extended to the velocity range $|\vlos|\!<\!50\kms$, consistently with the method used by \citet{Sembach+03} and \citet{Savage+03} to infer upper limits from the non-detections.
This maximizes the column density predicted, since in our model most of the gas occupies that range of velocities as shown in Fig.\,\ref{wake_lv}.

\begin{figure}
\begin{center}
\includegraphics[width=0.5\textwidth]{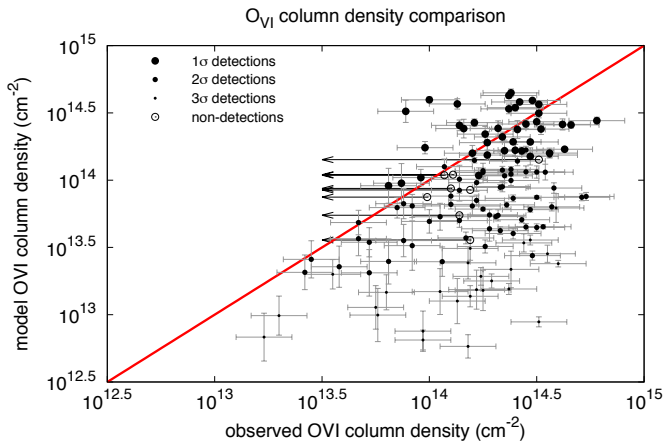}
\caption{Comparison between the \ovicap\ column densities observed (x-axis) and those predicted by our model of fountain wakes (y-axis) for all those detections located inside the $3\sigma$ contour level shown in Fig.\,\ref{wake_lv}. Filled circles represent detections, different sizes are used to indicate different confidence levels. Empty circles represent non-detections. The straight line represents the function $y\!=\!x$.}
\label{coldens}
\end{center}
\end{figure}

Fig.\,\ref{coldens} compares the observed and the predicted \ovi\ column densities for all those detections that are located inside the $3\sigma$ contour level of our all-sky model shown in Fig.\,\ref{wake_lv}.
The absence of points in the leftmost part of the plot is due to the \ovi\ detection limit, which is a few times $10^{13}\cmmq$\citep{Sembach+03}.
Error bars for the predicted column densities are derived by considering an average error of $\pm5\kms$ on the line-width $b_w$, while for the observed ones we assume a constant error of $0.13$ dex.
Both these errors are set accordingly to \citet{Sembach+03} and \citet{Savage+03}.
We point out that the error on the predicted column density is likely underestimated, since the integration limits in eq.\,(\ref{NOVI}) are fixed arbitrarily by analogy with a gaussian profile.
For most detections the column densities predicted are consistent with the observed ones within an order of magnitude, indicating that our model predicts the correct amount of hot gas in the Galactic halo.
Interestingly, the \ovi\ detections predicted with a higher confidence level (larger points in Fig.\,\ref{coldens}) are also those whose column densities are better reproduced by our model.
This is a remarkable result, since detections are classified within a given confidence level only accordingly to their kinematics (see Section \ref{dataanalysis}).

Considering together the datasets of \citet{Sembach+03} and \citet{Savage+03}, the number of lines of sight where no \ovi\ absorption is detected decreases dramatically.
Only $8$ non-detections are found, with the respective upper limits on the column density.
It is interesting that for all the non-detections our model predicts column densities smaller than those inferred by the upper limits (empty circles in Fig.\,\ref{coldens}).
Note also that non-detection does not necessarily imply a lack of gas at a given line of sight.
In fact, the warm-hot material in the turbulent wake of a fountain cloud is clumpy (see Fig.\,\ref{wake_hydro}), and there is the chance that a pencil-beamed observation passes through different wakes without intercepting any of the cloudlets. 
This is possible given that the number of wakes intercepted by a generic line of sight is not too large (see Section \ref{nwakes}).

Column density measurements for single warm detections are not reported in \citet{Lehner+12}, so we consider the \siiii\ average column density $\avg{\log N_{\rm Si\,III}}=13.42\pm0.21$ obtained by \citet{Shull+09} as a representative case.
The value predicted by our model can be derived using the same approach adopted for the \ovi\ component.
We consider the model of fountain wakes + outflows described in Section \ref{refinement} and we assume $\frac{n_{\rm Si_{III}}}{n_{\rm Si}}=0.903$ \citep[][for $\log(T)\!=\!4.45$]{SutherlandDopita93}, $\log[Z/Z_{\odot}]\!=\!-0.24$ (from the M11 simulation) and $\log(n_{\rm Si}/n_{\rm H})_{\odot}\!=\!-4.45$ \citep{AndersGrevesse89}. 
The integral in eq.\,(\ref{NOVI}) is extended to $70\kms$ around the velocity centroid of the line, consistently with the average line broadening found by \citet{Shull+09}.
The predicted \siiii\ column density averaged over all the absorbers is $\avg{\log N_{\rm Si\,III}}=13.44\pm0.36$, in remarkable agreement with the observed value \citep[see also][]{Fraternali+13}.

We stress again that these column densities are not derived by fitting our model to the \ovi\ or to the warm data, but they are instead fully determined by the fit to the \hi\ data of MFB12 and by the physical connection between the \hi\ phase and the warm-hot phase via the fountain-corona interaction.
Thus, this comparison strongly reinforces the validity of our scheme for the origin of the absorbers.

\subsection{The number of wakes per line of sight} \label{nwakes}
A further test of the validity of our model comes from the number of warm and \ovi\ absorptions observed for a given line of sight, which ranges from $0$ (non-detections) up to $4$ (some \ovi\ detections shown in Fig.\,\ref{wake_lv}).
If we average the number of \ovi\ detections combining the data of \citet{Sembach+03} and \citet{Savage+03}, we find a value of $1.6$ detections per line of sight. 
For the warm absorptions, the data of \citet{Lehner+12} give $0.7$ detections per line of sight, but this value is likely underestimated because the velocity range $|\vlos|\!<\!90\kms$ is excluded by the observations.
To investigate whether these numbers are consistent with the considered scenario of fountain wakes, we proceeded as follows.

Since the total gas mass of the Galactic halo is $3\times10^8\msun$ (MFB12) and the \hi\ mass of the clouds in the standard simulations of M11 is $2.4\times10^4\msun$ \citep[consistent with the typical \hi\ mass of an IVC, e.g.][]{vanWoerden+04}, we expect that the number of fountain clouds in the halo of the Milky Way is about $10^4$.
We consider the best model for the \hi\ halo of the Milky  Way obtained by MFB12, and we fit the resulting axi-symmetric \hi\ density distribution $\rho(R,z)$ with the following formula \citep[see][]{Marinacci+10b}:
\begin{equation}\label{halodistrib}
\rho(R,z) \propto \left(1+\frac{R}{R_g}\right)^\gamma \exp\left(-\frac{R}{R_g}\right)\exp\left(-\frac{z}{h_g(R)}\right)
\end{equation}
where
\begin{equation}\label{scaleheight}
h_g(R) = h_0 + \left(\frac{R}{h_R}\right)^\delta\ , \delta\ge0.
\end{equation}
The fit gives $R_g\!=\!1.08\kpc$, $h_0\!=\!175\pc$, $h_R\!=\!10.01\kpc$, $\gamma\!=\!6.30$ and $\delta\!=\!1.89$.
Then, we considered a three-dimensional cartesian space $(x,y,z)$, where the plane $(x,y)$ represents the Galactic midplane and the axis $z$ is the rotation axis of the Milky Way. 
In this space, we randomly generate $10^4$ particles with space-density distribution set accordingly to eq.\,(\ref{halodistrib}).
Each particle represents both the warm and the hot phase of a cloud's wake.
In Section \ref{effectofvrel} we show that the relative velocity between the clouds and the corona $v_{\rm rel}$ can increase up to $100\kms$ during the initial and the final part of the orbits.
Since $v_{\rm rel}$ largely affects the size of the wake left behind the cloud's path, we evaluate the volume occupied by the warm and the hot gas by analyzing both simulations at $v_{\rm rel}\!=\!75\kms$, shown in Fig.\,\ref{wake_hydro}, and $v_{\rm rel}\!=\!100\kms$. 
Typical volumes are calculated on the snapshot at $t=60\Myr$ by assuming a cylindrical symmetry along the major axis of the wake.
For $v_{\rm rel}\!=\!75\kms$, we found $3.8\times10^6\pc^3$ and $7.8\times10^6\pc^3$ for the warm and the hot component respectively.
By increasing the velocity to $100\kms$, the volume occupied by the warm (the hot) material increases by a factor $2.5$ ($9.5$), which implies that the angular filling-factor grows by a factor $1.8$ ($4.5$).
Finally, we consider an observer at coordinates $(x\!=\!0,y\!=\!8.3\kpc)$ scanning a large number of lines-of-sight in the simulated sky at $0\de\!\le\!l\!<\!360\de$ and $20\de\!\le\!|b|\!\le\!80\de$, consistently with the region where real absorptions are observed, and we count the number of warm and hot particles intercepted by each line of sight.

We find that the average number of particles intercepted ranges from $0.5$ to $1$ particles per line of sight for the warm wakes, and from $0.8$ to $4.0$ per line of sight for the hot wakes, depending on the sizes considered and on the different stochastic realizations of the particle distribution.
Despite the uncertainties, these values are fully consistent with the average numbers of warm and hot absorptions per line of sight found in the data.

\section{Discussion}\label{discussion}

\subsection{Limitations of the model}\label{limitations}
We have already pointed out that our model fully relies on the modelling of the Galactic \hi\ halo made by MFB12 and on the simulations of cloud-corona interaction of M11.
Both these works have their limitations, which in turn can affect the results presented in this paper.

The dynamical model of MFB12 is symmetric with respect to both the rotation axis and the midplane, thus all the features related to asymmetries in the \hi\ distribution or to specific \hi\ clouds are not taken into account.
Also, the parameters of the model (i.e. the mean kick velocities of the particles, the ionisation fraction and the coronal accretion rate) are kept constant in space.
This implies that the model can account only for the global properties of the Galactic \hi\ halo. 
The same is true for the warm-hot gas of our model of fountain wakes: non axisymmetries \citep[e.g. spiral arms,][]{StruckSmith09} or isolated HVCs can not be reproduced (but see Section \ref{HVCs}).

The main limitations of the M11's simulations are three: a) they are 2D; b) they are evolved up to $60\Myr$; c) gravity is not considered.
Turbulence, which is an important ingredient for our models, can change significantly when passing from 2D to 3D.
By increasing the amount of turbulence, the number of absorption features reproduced by our model increases accordingly.
Also, the typical orbital time for a particle in the dynamical model of MFB12 is $\sim80\Myr$, while the cloud-corona interaction in the simulations is followed only for $60\Myr$. 
Beyond that time, masses and velocities for the warm-hot particles are simply extrapolated from the simulation (see Section \ref{wake}). 
Large deviations from this extrapolated behavior could produce differences in our modelcubes of warm-hot gas.
Finally, the absence of gravity in the simulations implies that we must rely on the dynamical model of MFB12  to infer the orbits of the warm-hot gas in the wakes.
However, it is possible that such material does not follow exactly the same orbit of the cold gas, or that the velocity gap between the different phases is larger than that determined in Section \ref{wake}. 
This latter effect, if present, may produce a warm-hot medium that rotates more slowly than in the present model. 
This would help to reproduce some of the highest-velocity absorption features.

A further source of uncertainties is given by the initial conditions of the simulations: they start considering a spherical cloud of cold gas, while the gas ejected by a supershell/bubble would be already fragmented and (partially) ionised from the beginning \citep[see][]{Melioli+08}.
A way to solve this discrepancy is to include in the model the ionised outflowing part of the fountain clouds, as we did in Section \ref{refinement}.
The warm particles escaping from the disc during the first $15-20\Myr$ of the orbit, together with those associated with the fountain-cloud wakes, provide a better description of the warm absorption dataset.
All these uncertainties considered, it is remarkable that a simple, axi-symmetric model built specifically to fit the \hi\ data reproduces both the kinematics and the column densities of such a large fraction of warm and hot absorption features in the halo of the Milky Way.

\subsection{The effect of the cloud-corona relative motion}\label{effectofvrel}
The outcome of the M11's simulations depends on the relative velocity $v_{\rm rel}$ between the cloud and the corona, which M11 found to spin with a given velocity lag with respect to the cold material.
In our analysis we have fixed $v_{\rm rel}$ to $75\kms$, which is a representative value of the lag in the azimuthal direction between the two media.
However, because of the presence of non-circular (vertical+radial) motions during the beginning and the final part of the cloud's orbit, $v_{\rm rel}$ varies with time as:
\begin{equation}\label{vrel}
v_{\rm rel}(t) = \sqrt{v^2_z(t) + \Delta v_{\phi}^2 + v^2_R(t)}
\end{equation}
where $\Delta v_{\phi}\!=\!75\kms$ is the azimuthal lag, while $v_z$ and $v_R$ are respectively the vertical (i.e. perpendicular to the disc) and the radial (i.e. perpendicular to the rotation axis) component of the cloud's velocity.
Since, on average, clouds are ejected from the disc with a vertical speed of $70\kms$ (MFB12),  $v_z$ and $v_R$ are not negligible during the initial and the final part of the orbits and $v_{\rm rel}$ can increase to a maximum value of $\sim100\kms$.
We tested the impact of this larger speed by considering the extreme case of a constant lag between the clouds and the corona of $100\kms$.
For this purpose, we considered the hydrodynamical simulation of M11 at $v_{\rm rel}=100\kms$ and we repeated the analysis previously done for the case at $v_{\rm rel}=75\kms$.
We found that the fraction of warm and \ovi\ absorption reproduced by this new model is respectively $85.2^{+5.0}_{-2.4}\%$ (without considering outflows) and $48.0^{+2.5}_{-1.2}\%$, not very different from those found for the case at $v_{\rm rel}=75\kms$.
This indicates that the change of $v_{\rm rel}$ during the orbit of the fountain clouds does not play a major role in shaping the position-velocity distribution of the warm-hot gas.
The impact of a larger $v_{\rm rel}$ is instead particularly significant for the sizes of the fountain-cloud wakes, as we discussed in Section \ref{nwakes}.

\subsection{Relation to the classical HVCs} \label{HVCs}
Even though the low metallicity of the classical HVCs points firmly towards an extragalactic origin, there is no general consensus on how these systems are generated.
The position-velocity plots of Fig.\,\ref{wake_lv} shows a significant overlapping between the HVCs and the warm-hot gas of our model.
Gas at such high velocities was not predicted in previous models of the galactic fountain \citep[e.g.][]{HouckBregman90}, and in our scheme is the result of the turbulent mixing between the cold gas of the fountain clouds and the hot gas of the corona. This process increases the velocity dispersion of the warm-hot gas composing the wake of the fountain clouds up to values of $\sim30\!-\!50\kms$, so that this material is distributed over a range in velocities that encompasses that of the classical \hi\ HVC complexes.
But can the galactic fountain produce the classical HVCs?
In our scheme, a direct connection between the fountain mechanism and these systems seems quite unlikely, because the latter are located at distances above the plane that are not reached by the particles ejected from the disc.
In fact, the scale-height of the hot material in the fountain-cloud wakes is only slightly larger than that found by MFB12 for the \hi\ phase of the Galactic halo, being $\sim1.2\kpc$ at $R\!=\!R_{\odot}$ and increasing up to $3\!-\!4\kpc$ at $R\!=\!12\kpc$.
In order to bring cold particles at line-of-sight velocities, Galactocentric radii and heights consistent with those of - for instance - the complex C, our model requires ejection velocities $\gtrsim175\kms$ occurring at the periphery of the stellar disc.
Given that the \hi\ mass of the complex C is $5\times10^6\msun$ \citep{Thom+08}, the energy associated to the ejection would be $\gtrsim3\times10^{54}$ erg, corresponding to $\sim30000$ simultaneous supernova explosions (considering $10^{51}$ erg per SN and an efficiency of $10\%$) which is a very unlikely event.
Our values of the scale-height of the hot material are compatible with existing estimates from \ovi, for instance \citet{Zsargo+03} found $2.3-4.0\kpc$ and \citet{Bowen+08} found $3.2-4.6\kpc$. However this comparison is not straightforward, because these estimates are derived by assuming an exponential profile with constant scale-height and midplane density for the \ovi\ within a few kpc from the Sun, while our model predicts strong variations of both parameters.

Fig.\,\ref{wake_lv} shows also that the position-velocity locations of several warm and hot detections overlap with those of the HVCs.
\citet{Lehner+12} have argued that all the detected warm absorption features originate in the ionised envelope that surround these complexes \citep[e.g.][]{Sembach+99}. 
This would place the absorbers at typical distances of $\sim3-9\kpc$ from the Galactic disc \citep{LH11}, while in our scheme the warm-hot gas is located only within a few kpc from the midplane (MFB12).
Information on the metallicity cannot be used to discriminate between the two origins, given that a) the warm gas of our model has metallicities ranging from $\sim0.2$ solar to solar; and b) the value that is commonly assumed for the high-velocity gas in the Galaxy ($0.2\zsun$) has been determined specifically for complex C \citep[e.g.][]{Wakker+99,Collins+03,Tripp+03}, and it is unclear to what extent this is representative of the whole population of absorbers.
In the following, we check how our results change if we consider that some absorptions are related to the classical HVCs.

We isolated the \hi\ emission of the \hi\ HVCs in the LAB datacube, and we excluded all those detections that are separated by less than $10\de$ in the angular direction and by less than $60-100\kms$ in the velocity direction - depending on whether we consider the warm or the \ovi\ detections - from the $2\sigma_{\rm rms}$ contour of the \hi\ complexes in the LAB survey.
The number of warm and \ovi\ absorptions decreases respectively from 72 to 31 and from 159 to 138.
We re-evaluated the fraction of absorptions reproduced by the model (without considering outflows), finding a modest increment to $60.1^{+3.0}_{-1.4}\%$ in the \ovi\ case, while for the warm detections the fraction decreases to $37.8^{+6.4}_{-3.0}\%$.
However, this value is biased by the fact that most HVCs are located at negative velocities, thus, after the cut, the warm dataset is dominated by detections at $\vlos\!>\!0$.
Using the model with the outflows, virtually all the warm absorptions are reproduced ($100^{+0}_{-4.1}\%$), consistently with the previous results (Section \ref{refinement}).
Even though including or not these detections does not affect our results significantly, we can not exclude that a fraction of the warm absorptions of \citet{Lehner+12} is generated in the ionised envelope of the HVCs.
The mechanism producing the warm-hot gas around these large complexes is likely to be the same as that producing the wakes of the fountain clouds, although the cooling of the corona may be less efficient given the low metallicity of the HVCs and the lower density of the corona at their typical distances.
Photoionisation from both Galactic and extragalactic UV background can also play a role.

\subsection{The high-velocity \ovi\ absorbers}\label{missingdetections}
Our model of fountain wakes does not reproduce a significant fraction ($\sim45\%$) of the \ovi\ absorptions.
Some of these features could be probably reproduced by tuning some parameters in the dynamical model of MFB12 and in the simulations of M11.
However, most of them occur at velocities that are largely inconsistent with those predicted by our model (see, for instance, the top-right panel of Fig.\,\ref{wake_lv}).
If we limit our analysis to the 68 high-velocity ($|\vlos|\!>\!90\kms$) \ovi\ detections of \citet{Sembach+03}, the fraction reproduced by our model decreases down to $34.1^{+3.7}_{-1.8}\%$.
Here, we discuss two possible different origins for these high-velocity \ovi\ absorption features.

\begin{figure}
\begin{center}
\includegraphics[width=0.5\textwidth]{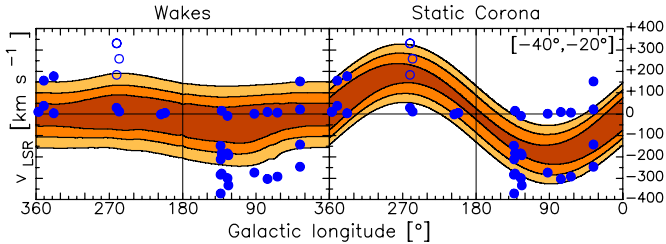}
\caption{Longitude-velocity diagram at $-40\de\!\le\!b\!\le\!-20\de$ for the hot gas component of our model of fountain wakes (\emph{left panel}) and for a static corona with density distribution $\propto r^{-3/2}$ (\emph{right panel}). Filled circles represent \ovicap\ detections, empty circles represent detections related to the Magellanic system. Contour levels from the innermost to the outermost at $1$, $2$ and $3\sigma$ (see Section \ref{dataanalysis}).}
\label{corona}
\end{center}
\end{figure}

A first possibility is that these absorptions occur in the circumgalactic medium of the Milky Way, far beyond the regions where the galactic fountain mechanism takes place.
\citet{Zsargo+03} studied the \ovi\ absorption features towards a sample of 22 halo stars located between 3 and 5 kpc from the Galactic plane, finding that all the detections occur at $|\vlos|\!<\!100\kms$.
This would imply that the \ovi\ absorbers at larger velocities found by \citet{Sembach+03} are likely to be located in the outskirt of the Galactic corona.
\ovi\ absorptions at large distances from the hosting galaxies are commonly observed up to redshift $\sim3$ \citep{Fox11}.
These detections have some peculiar properties: they are always located within half a Mpc of projected distance from a galaxy with luminosity $L\!>\!0.25 L_*$ \citep{WakkerSavage09}, their column density is always around $10^{14}\cmmq$ \citep{Fox11}, and to about half of them a Ly$\alpha$ absorption line can be associated within $50-200\kms$ \citep{Stocke+06}.
The origin of the absorber material is controversial: UV background photoionisation of cosmological filaments, strong outflows from the hosting galaxy and cooling flows from the circumgalactic medium are all viable possibilities.
Thus, it is possible that a fraction of the high-velocity detections belongs to a second population of absorbers, placed in the outskirt of the Galactic corona and completely unrelated to the cooling mechanism proposed here.
To test this hypothesis, we built a simple model of a \emph{static} corona with density profile $\propto r^{-3/2}$ \citep{FukugitaPeebles06}, and we compared the resulting modelcube with the \ovi\ observations via the technique described in Section \ref{dataanalysis}.
We assumed for the coronal material a velocity dispersion similar to that of the hot gas in the wakes of the fountain clouds ($49\kms$).
Fig.\,\ref{corona} compares the l-v diagram around $b\!=\!-30\de$ for our model of fountain wakes with that derived for this static coronal medium.
Clearly, the latter allows to reproduce detections with a large $\vlos$, even if it can not explain most of the \ovi\ detections at $|\vlos|\!<\!90\kms$.
The fraction of high-velocity \ovi\ detections reproduced by this new model is $62.8^{+4.6}_{-2.2}\%$.
This result is consistent with that of \citet{Sembach+03}, who found that the high-velocity \ovi\ features are more clustered about the velocity of the Galactic Centre.
This simple test suggests that a significant - but not dominant - fraction of \ovi\ absorbers analysed in this work may take place in the outskirt of the Galactic corona, at tens or even hundreds of kpc from the gaseous thick disc of the Milky Way.

\begin{figure}
\begin{center}
\includegraphics[width=0.5\textwidth]{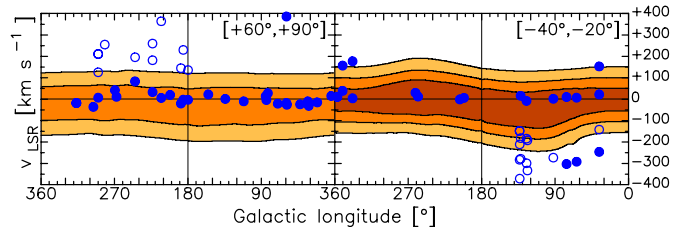}
\caption{Longitude-velocity diagrams at $+60\de\!\le\!b\!\le\!+90\de$ (\emph{left panel}) and $-40\de\!\le\!b\!\le\!-20\de$ (\emph{right panel}) for the hot gas component of our model of fountain wakes. Filled circles represent \ovicap\ detections that we consider related to the Galactic halo, while empty circles represent \ovicap\ detections that can be related to nearby galaxies (see text). Detections related to the Magellanic system are not shown. Contour levels from the innermost to the outermost at $1$, $2$ and $3\sigma$ (see Section \ref{dataanalysis}).}
\label{noGAL}
\end{center}
\end{figure}

Another possibility is that part of the \ovi\ detections are not related with the Milky Way, but they occur in the circumgalactic medium of external galaxies.
As a test, we retrieved from the Nasa/Ipac Extragalactic Database a list of UGC and NGC galaxies with systemic velocity smaller than $600\kms$.
Then we used the HyperLeda catalogue \citep{Paturel+03} to obtain distances ($d$), inclination angles ($i$) and rotation velocities ($v_{\rm rot}$) for all the galaxies in our sample.
When these data were not available, we adopted average values based on the other galaxies in the sample.
Assuming for the absorber material a velocity dispersion $\sigma$ of $49\kms$ and a typical distance to the galaxy centres $d_{\rm abs}$ of 500 kpc \citep{WakkerSavage09}, we removed from the dataset of \citet{Sembach+03} all those detections located at angular separation smaller than $\tan^{-1}(d_{\rm abs}/d)$ and velocity separation smaller than $v_{\rm rot}\sin(i)+2\sigma$ from the galaxies in our sample.
After this cut, the number of high-velocity \ovi\ features decreases significantly from 68 to 27.
Fig.\,\ref{noGAL} shows that most of the \ovi\ absorptions visible at positive velocities at $+60\de\!\le\!b\!\le\!+90\de$ are removed, as well as those located at $-40\de\!\le\!b\!\le\!-20\de$ and $l\!\simeq\!130\de$ in a stripe at negative velocities.
These latter are absorptions related to M31.
Our model of hot fountain wakes reproduces $65.2^{+8.3}_{-3.9}\%$ of the remaining 27 detections.
When including also the detections at $|\vlos|\!<\!90\kms$, the fraction reproduced becomes $71.7^{+3.5}_{-1.7}\%$, $\sim17\%$ larger than that estimated in Section \ref{results}.
It is interesting to note that if we assumed $d_{\rm abs}\!=\!50\kpc$ only 4 \ovi\ features would be removed. This indicates that these absorption lines must be associated to extended media surrounding the nearby galaxies rather than to the galaxy discs.
We conclude that some percentage of the \ovi\ absorbers analysed in this work may be located outside the virial radius of the Milky Way and produced in the circumgalactic medium of nearby disc galaxies.

\subsection{Gas accretion onto the disc}\label{gasaccretion}
The ionised gas surrounding our Galaxy is regarded as one of the main candidates for replacing the material consumed by the process of star formation in the disc.
\citet{LH11} used all the information available on their warm absorptions (e.g. upper limits on distances, sky covering fraction, ion column densities) to infer a mass accretion rate onto the disc of $0.8-1.4\msunyr$. 
This value was obtained by considering an average infall velocity of $90-150\kms$.
Using similar arguments but different absorption data, \citet{Shull+09} inferred an accretion rate onto the disc of $\sim1\msunyr$.
These values are derived regardless of the phase of the gas involved, and it is not clear whether the warm material cools during its infall or remains ionised.

In our model, because of the coronal condensation, the fountain cycle returns to the disc more material than what has been ejected from it by stellar feedback.
A pristine, multi-phase gas accretion is thus `hidden' in the fountain cycle.
The cold phase of the accreting material can be inferred via the fitting of the \hi\ kinematics: the dynamical model of MFB12 suggests a global accretion of neutral gas (H+He) onto the disc at a rate of $\sim2\msunyr$.
In addiction, a `warm' accretion is provided by coronal material that, though entrained by the clouds' wakes, is not able to cool down to recombination temperatures before impacting with the disc.
We used the model of MFB12 to infer the accretion rate of this warm gas, obtaining a value of $\sim1\msunyr$ after correcting for the He abundance.
This value is perfectly consistent with that estimated in the works mentioned above for the ionised gas. 
What remains unclear is whether this ionised material can take part in the process of star formation in the Galactic disc.

\section{Conclusions}\label{conclusions}
Accretion of pristine gas from cosmological coronae seems to be required in order to explain the properties and the evolution of disc galaxies.
Hydrodynamical simulations indicate that the cooling of the corona can be triggered by the passage of metal-rich gas clouds, like those ejected in star forming galaxies by supernova feedback (galactic fountains).
The material cooling from the corona is expected to trail the cloud in an extended turbulent wake.
In this framework, galaxy discs are surrounded by thick layers of multiphase gas at temperatures below the virial temperature produced by the interaction of the fountain with the coronal medium.

In this paper we tested the above scenario for the Milky Way.
We studied whether the warm-hot gas present in the wakes of the fountain clouds is responsible for the ion absorption lines (\ovi, \siiii, \siiv, \cii, \civ) that have been observed towards several background sources.
We modelled the all-sky position-velocity distribution of this gas by combining the galactic fountain \hi\ model of MFB12 with the hydrodynamical simulations of M11.
Our model does not have free parameters as it is fully determined by the fit on the \hi\ component of the Galactic halo performed by MFB12.
By comparing the properties of the absorptions detected by \citet{Sembach+03}, \citet{Savage+03} and \citet{Lehner+12} with those predicted for the warm-hot gas by our model, we obtained the following results:
\begin{enumerate}
\item the great majority ($75\!-\!95\%$) of the `warm' absorptions (\siiii, \siiv, \cii, \civ) and about half of the \ovi\ absorptions have a position-velocity distribution fully consistent with that predicted by our model;
\item the \ovi\ and the \siiii\ column densities of the absorbers agree with those predicted by our model;
\item the large velocity spread of the absorbers is consistent with the predictions of hydrodynamical simulations and it is due to the development of turbulence in the clouds' wakes;
\item the average number of absorptions detected per line of sight is consistent with the average number of wakes intercepted by the line of sight in our model;
\item excluding from our analysis all those absorptions that can be associated to the classical HVCs does not change significantly the above results.
\end{enumerate}

We conclude that half of the \ovi\ absorbers and virtually all the warm absorbers are produced within a few kiloparsecs above the disc, at the interface between the latter and the cosmological corona of the Milky Way. 
The material cooling from the corona provides an accretion of pristine gas onto the disc at a rate of $\sim2\!-\!3\msunyr$.
This coronal accretion is an adequate supply of material to maintain star formation in the Galaxy disc.
\citet{Fraternali+13} suggests that this is the main mode by which the Milky Way is accreting gas from the environment at the current epoch and may explain how the hot mode of cosmological accretion feeds star formation in disc galaxies.

Half of the \ovi\ detections are not reproduced by our model.
We speculate that they constitute a different population of absorbers, similar to those observed in extragalactic surveys. 
These features do not seem to participate in Galactic rotation and are likely located in the circumgalactic medium of the Milky Way and/or of nearby disc galaxies, at distances of tens or hundreds of kpc from the discs.   

\section*{Acknowledgments}
The authors are grateful to James Binney and an anonymous referee for providing insightful comments.
We also thank Carlo Nipoti for helpful comments.
We aknowledge financial support from PRIN MIUR 2010-2011, prot.\ 2010LY5N2T.
FM acknowledges support from the collaborative research centre ``The Miky Way system'' (SFB 881) of the DFG, and the CINECA Award N. HP10CINIJ0, 2010 for the availability of high performance computing resources.
\bibliographystyle{mn2e}
\bibliography{wakes}{}
\bsp

\label{lastpage}
\end{document}